# Avatar Appearance and Behavior of Potential Harassers Affect Users' Perceptions and Response Strategies in Social Virtual Reality (VR): A Mixed-Methods Study

XUETONG WANG*, The Hong Kong University of Science and Technology, China
ZIYAN WANG*, The Hong Kong University of Science and Technology (Guangzhou), China
MINGMIN ZHANG, Law School of Southwest University, China
KANGYOU YU, University of California, Santa Barbara, USA
PAN HUI†, The Hong Kong University of Science and Technology (Guangzhou), China
MINGMING FAN‡, The Hong Kong University of Science and Technology (Guangzhou), China

Sexual harassment has been recognized as a significant social issue. In recent years, the emergence of harassment in social virtual reality (VR) has become an important and urgent research topic. We employed a mixed-methods approach by conducting online surveys with VR users ($N = 166$) and semi-structured interviews with social VR users ($N = 18$) to investigate how users perceive sexual harassment in social VR, focusing on the influence of *avatar appearance*. Moreover, we derived users' response strategies to sexual harassment and gained insights on platform regulation. This study contributes to the research on sexual harassment in social VR by examining the moderating effect of avatar appearance on user perception of sexual harassment and uncovering the underlying reasons behind response strategies. Moreover, it presents novel prospects and challenges in platform design and regulation domains.

**ACM Reference Format:**
Xuetong WANG, Ziyan WANG, Mingmin Zhang, Kangyou Yu, Pan Hui, and Mingming Fan. 2024. Avatar Appearance and Behavior of Potential Harassers Affect Users' Perceptions and Response Strategies in Social Virtual Reality (VR): A Mixed-Methods Study. In *Proceedings of The 27th ACM SIGCHI Conference on Computer-Supported Cooperative Work Social Computing (CSCW) (CSCW '2024)*. ACM, New York, NY, USA, 27 pages. https://doi.org/10.1145/3686934

## 1 Introduction

Online sexual harassment has been reported in many online platforms, including social networking sites [60, 64], online games [12, 24, 56], and even in virtual reality (VR) environments [3, 63] in

---

*Both authors contributed equally to this research.
†Pan Hui is also affiliated with Hong Kong University of Science and Technology, and University of Helsinki, Finland
‡Corresponding author

---

Authors' Contact Information: Xuetong WANG, xwangdd@connect.ust.hk, The Hong Kong University of Science and Technology, Hong Kong, China; Ziyan WANG, webmaster@marysville-ohio.com, The Hong Kong University of Science and Technology (Guangzhou), Guangzhou, China; Mingmin Zhang, Law School of Southwest University, Chongqiong, China, zhangmingmin1998@163.com; Kangyou Yu, University of California, Santa Barbara, California, USA, kangyouyu@ucsb.edu; Pan Hui, The Hong Kong University of Science and Technology (Guangzhou), Guangzhou, China, panhui@ust.hk; Mingming Fan, The Hong Kong University of Science and Technology (Guangzhou), Guangzhou, China, mingmingfan@ust.hk.

---

recent years. Due to the anonymity of social networks, sexual harassment easily and frequently happens on social platforms, which is considered a serious social issue due to its negative effect on victims [1, 36]. As one of the newly emerging technologies and social platforms, VR has become increasingly popular with the public. An average of one accident of abuse and harassment per 7 minutes is reported in VR Chat, one of the most popular social VR apps [22]. Because of the immersion and multimodal interaction of VR, users may have a higher level of perception of sexual harassment in social VR platforms than in common online platforms. Consequently, VR harassment has a higher possibility of happening to users in social VR platforms [4]. Those phenomena underscore the urgent need to combat the rising harassment in VR.

Nevertheless, the unique characteristics of VR environments pose new challenges to regulating sexual harassment in VR. Specifically, the interactive and multi-modal nature of VR systems enables a broader range of harassment behaviors, such as body groping, that are impossible in traditional online social media settings. This evolution has led to constant debate topics such as whether groping in VR is actually groping [3]. In a broader sense, it remains unclear which behaviors are considered sexual harassment in VR, and if applicable, whether and how they be responded to and regulated. To fill in this gap, recent research began to study and clarify sexual harassment in VR for platform governance. For example, Blackwell et al. classified harassment in social VR platforms into verbal, physical, and environmental harassment [5]. Another line of research [28, 52] contributes qualitative interview studies that explore VR harassment experiences to conceptualize VR harassment. Although some studies focused on sexual harassment in social VR and the effect of avatars on users' perception, little is known about the impact of avatar appearance and behaviors on users' perception of harassment in social VR platforms. Moreover, previous studies mainly leveraged qualitative research on relatively small sample sizes and focused on the experience instead of the social acceptance, responses, and regulatory measures of VR harassment. We argue that surveying a larger scale of participants would contribute to a more comprehensive understanding of public opinions toward VR harassment, which is essential to establishing social norms for regulating the VR environment.

In addition, cultural background is an essential factor to consider in research on online communities, as users from different cultural backgrounds may exhibit distinct online social behaviors and perceptions [68]. Current studies on harassment in VR social platforms have predominantly focused on users from Western cultural backgrounds. To address this gap and enhance the diversity of research backgrounds, this study targets Chinese users as research subjects and distributes an online survey to public forums to get a large number of data of more comprehensive and quantitative opinions of VR harassment from the public before understanding their responses in depth via semi-structured interviews.

Specifically, we report our findings of a mixed-methods study, which includes a survey and an interview, to answer the following research questions:

- **RQ1:** How do VR users perceive harassment in terms of potential harasser's avatar appearance and behavior?
- **RQ2:** How do VR users respond to potential harassment and what are their opinions of platform governance?

Our research contributes to a gap in the effect of harassers' avatar appearance and behavior on users' perception of harassment in social VR. Firstly, we present an online survey in a Chinese forum asking for public opinions about perception and response to harassment in 25 different scenarios that we created based on the literature. We received a total of 166 valid responses and found that there was a general consensus among the public regarding the judgment of harassment behaviors. However, the interpretation and perception of harassment were found to be influenced by avatar appearance. Regarding their response strategies to potential harassment behaviors, participants tended to adopt non-confrontational response strategies. Secondly, we conducted in-depth semi-structured interviews with volunteers from the online survey. We conducted a detailed exploration of how appearance specifically influences user perception and the reasons behind users' adoption of non-confrontational responses. Additionally, we also addressed the important issue of platform regulation in our interviews with the participants. Finally, based on the findings, we discussed potential reasons behind these phenomena and put forth design recommendations for platform regulation. In summary, our contributions are as follows:

- We conducted a mixed-methods study to study public opinions towards the perception and response strategies of sexual harassment in social VR, including an online qualitative survey with 166 participants and semi-structured interviews with 18 interviewees;
- We investigated the impact of user gender, harasser's avatar appearance, and harasser's behavior on user perception and uncovered potential reasons behind it;
- We explored users' tendencies in responding to sexual harassment and highlighted potential challenges and opportunities with platform regulation in social VR.

## 2 Related Work

Our work is motivated and informed by prior work on Social VR, Online sexual harassment in the VR platform, Self-embodiment in online space, and the Governance of social VR platforms.

### 2.1 Sexual Harassment in VR

In the past few years, with the popularity of VR HMDs and the immersive experience, emerging social VR applications also boomed rapidly (e.g., VR Chat, Altspace VR, RECroom, and so on). In the traditional 3D video game and virtual world, users are required to sit in front of a computer and look at a screen to manipulate the avatar. Social VR provides more possibilities for users to control the avatar and is more flexible for users to control the avatar to represent themselves by using the whole-body movement tracking.

Compared with the previous online social interaction, which is limited to text, voice, and image communication, social VR could provide a wider range of interaction methods which includes physical interaction and verbal interaction. These features not only bring a more immersive experience but also increase potential harassment risk [4]. Potential harassment stems from the presence, body tracking, synchronous voice conversations, and simulated touching and grabbing capabilities provided by VR technology[5, 25, 83]. In addition, the sense of presence brought by VR technology will also exacerbate the discomfort of users facing harassment [50].

In the real world, there is a system of universal values in society, which will consider harassment to be bad, incorrect, unethical, or even illegal. But in the virtual world, there is often a new system of reconstructed cultural and moral standards. In some toxic gaming cultures, people who take insulting actions are considered more masculine [11]. Those who react strongly to harassment will instead be perceived as being too serious and emotional and thus unpopular with other players. This phenomenon is more obvious in competitive games. This bad culture will prompt players to use provocative language and make insulting actions to gain a sense of group identity [4].

There have been many studies investigating VR harassment. A survey of frequent users of HTC Vive, Oculus Rift, PlayStation VR, and Microsoft Windows Mixed Reality found that 49% of women and 36% of men in the survey group experienced harassment [62]. A previous survey by Shriram and Schwartz showed that 42% of respondents said they had witnessed VR harassment [77]. Results show that vulnerable groups such as women, transgender people and people of color are more likely to suffer from VR harassment [13, 19, 24, 48, 56, 67, 72, 76, 82], studies conducted on online gaming and virtual worlds also arrive at comparable findings [8, 13, 33, 69, 80]. The research and governance regarding sexual harassment behaviors on social VR platforms are imperative.

## 2.2 Self-embodiment in online space

In the real world, we usually perceive ourselves through interaction with the environment, and our real body is the medium of this interaction. In the virtual environment, everything is virtual, and our physical body cannot really enter the virtual world to interact with the virtual environment. In this case, the avatar becomes the carrier of communication.

In previous studies on self-awareness and cognition, it was pointed out that the body has two forms: online presentation and offline presentation [9]. Online presentation refers to the current representation of one's body, while offline presentation refers to the expected appearance in the long term. Inconsistency between online and offline body representations can result in the phenomenon of "phantom limb," where users still perceive the presence of a lost limb. The "rubber hand" experiment, as demonstrated in [7, 41], illustrates a similar concept of the three-way interaction between touch, vision, and proprioception. This phenomenon has also been confirmed in virtual environments [87]. Thus, the inconsistency between the avatar appearance and the user, results in a blurred and complicated definition of sexual harassment in the virtual environment becomes blurred and complicated. Certain actions that are clearly offensive in the physical world may be acceptable in the virtual environment. Conversely, common behaviors in the real world may cause discomfort in a virtual environment.

Since avatars are projections of real users, some studies have mentioned that one's avatar may have social consequences, and avatars can provide some personality information at the level of personal characteristics [21]. Therefore, the image of the avatar has an impact on the user's perception and judgment of behavior.

Previous studies on social VR have focused on the personal characteristics of the victimized party. For example, women and marginalized people are more likely to be harassed [28], and female users are more sensitive to sexual harassment [84]. However, they have not yet summarized the impact of the avatar appearance of the sexual harassment implementer on behavior perception and Judgmental impact.

In prior studies on sexual harassment, it has been demonstrated that the gender of the harasser can influence the perception and judgment of the harassed individual towards the behavior. Harush and Kaspi-Baruch compared the impact of harassers of different genders on observer perception among Israeli users in ambiguous situations occurring on WhatsApp [66]. LaRocca and Kromrey investigated the manipulation of the gender and physical attractiveness levels of the perpetrator/victim, within a sample of university students [45]. Cummings and Armenta examined peer sexual harassment in an academic context, exploring the influence of the harasser's gender, participant gender, severity of harassment, and the presence of bystanders [14]. These studies converged on a similar conclusion: when engaging in similar behaviors, the actions of male harassers are more frequently interpreted as sexual harassment. In the context of social VR platforms, harassers communicate through their avatars. Therefore, we speculate that behaviors exhibited by users utilizing male avatars are more likely to be perceived as harassment compared to those exhibited by users employing female avatars.

In addition, nearly all social VR platforms support user-customized avatars, particularly VRChat, which allows users to upload their own avatars, providing them with a high degree of freedom. User avatars can extend beyond the limitations of being 'human'. Previous research has extensively investigated the impact of realism and anthropomorphism on avatar appearance within immersive applications. Latoschik et al. indicated that highly anthropomorphic avatars are perceived as more human-like choices and are more easily accepted in terms of virtual body ownership, but they may trigger the uncanny valley effect when used by others [46]. Girondini et al. found that more realistic virtual audience avatars and voices in virtual reality public speaking scenarios led to higher levels of anxiety [31]. Another study [40] focused on augmented reality remote collaboration applications, where collaborators using realistic full-body avatars were considered the optimal choice for remote collaboration, while cartoon styles were seen as stylistic preferences. Yuan et al.'s research revealed that in virtual reality interview tasks, participants generally found realistic avatars to be more trustworthy and affable [86]. These studies indicate that in immersive applications, the level of anthropomorphism and realism in avatars potentially affects the sense of immersion, embodiment, and perception. While the impacts may vary in different scenarios, a common finding across these researches is that realistic avatars are generally perceived as more similar to humans in virtual environments compared to cartoon avatars.

Prior researches highlighted the impact of anthropomorphism on the attractiveness of avatars [32, 55, 75] and its significant influence on users' social presence and copresence [29]. Researchers emphasized the significance of considering the visual fidelity of virtual characters during their design, as it directly influences users' behaviors and interactions with them [17]. Thus, it remains unclear whether the level of anthropomorphism in avatars affects how users perceive and respond to harassment in social VR, warranting further exploration.

Based on previous research, we hypothesize that on social VR platforms, the gender of an avatar and its level of anthropomorphism may influence users' perception of harassment behavior. For example, behaviors exhibited by avatars with male features or a higher degree of anthropomorphism may be more likely to be defined as harassment. However, it appears that the current research is lacking in this particular aspect. To address this research gap, our study aims to categorize avatar types by their gender and level of anthropomorphism in order to understand the impact of the harasser's avatar on users' perception and judgment of behaviors related to sexual harassment.

The existing literature demonstrates that preventing harassment in social virtual reality (VR) is a crucial issue and that users' avatars play a decisive role in self-presentation and immersive interaction. Our study aims to investigate how users perceive sexual harassment behavior on social VR platforms and how various factors, particularly avatars, affect their perception levels (**RQ1**). The answers to this question can further promote accurate monitoring and prevention of sexual harassment behavior within the social VR community. After examining users' perception mechanisms of sexual harassment behavior, we are also motivated to explore how users respond to such behavior (**RQ2**) to understand the current problems on the platform and how to optimize them.

## 3 Survey

We employed a web-based questionnaire survey to reach a broader set of users of VR in China. This method allowed us to get a diverse sample of respondents at different stages of their genders and ages.

### 3.1 Survey design

The survey was separated into two sections. The first section consisted of four questions aimed at gathering respondents' demographic data, including age, gender, VR experience, and VR applications they had used. To ensure appropriateness, respondents were required to have reached the legal age of majority in China (Age≥18) due to the inclusion of scenarios with sexual implications. Furthermore, respondents were asked to have prior VR experience, as we believe that such individuals possess a fundamental understanding of VR immersion, and every VR user may be a potential user of social VR. To select representative VR applications, we referred to the data from *Steam* (one of the largest VR application platforms in China), considering the downloads of various application types. Additionally, we provided an "other" text box for inputting additional answers.

The second section comprised 25 scenarios (Table 1), each containing four questions. These scenarios were designed from two key aspects: avatar appearance and behavior (i.e., actions or activities being performed).

*3.1.1 Avatar Appearance Selection.* In this section, we elaborate on the process of selecting avatar appearances in the survey, taking into account various genders and degrees of anthropomorphism.

**Gender.** Gender cannot be simply ascertained by a singular visual element, presenting a considerable challenge in categorizing all genders solely based on appearance [27, 70]. By contrast, users more readily identified binary gender in an online survey. Therefore, this survey distinctly centers on exploring the impact of the male and female genders.

**Anthropomorphism.** In Nowak and Rauh's study [61], the authors discussed the definition of anthropomorphism from two perspectives: behavior and appearance. In the context of social VR, behavior is user-controlled, leading us to exclude behavior as a factor in the selection of avatars. Regarding appearance, the author highlighted that anthropomorphism can be defined as an object possessing human morphology or visual characteristics. Lugrin et al. categorized visual characteristics into two distinct groups: *anatomy* and *composition* [51]. "Anatomy" refers to the fundamental structural information of an object, such as its body parts and interconnections, and "composition" represents the attributes related to the shape, size, texture, and surface topology of these body parts. In this study, the authors differentiated avatars based on composition - the degree of simplification of human features, categorizing them into highly anthropomorphic avatars and abstract/stylized/cartoon humanoid avatars, which represent a human avatar that is genderless, race-less, and ethnicity-less. We adopt this classification method in our avatar selection process.

Based on this criterion, we selected several groups of similar avatars from the VRchat legacy avatar list, categorized according to the level of anthropomorphism. We chose legacy avatars because they are officially provided, more accessible, and align with the platform's official definition. In the "*low anthropomorphism (non-human)*" category, the avatars do not possess gender features. However, in the "*medium anthropomorphism*" and "*high anthropomorphism*" categories, the avatars do exhibit gender features. As a result, the latter two categories are further classified based on gender, resulting in the formation of five avatar categories: *High anthropomorphic male (Male_High)*, *High anthropomorphic female (Female_High)*, *Medium anthropomorphic male (Male_Medium)*, *Medium anthropomorphic female (Female_Medium)*, and *low anthropomorphic genderless (Genderless_Low)*. Finally, researchers randomly selected one avatar from each of the five categories for the experiment (Figure 1).

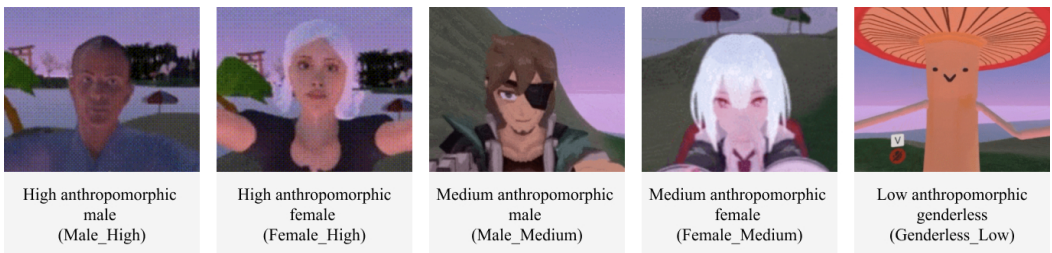

| High anthropomorphic male (Male_High) | High anthropomorphic female (Female_High) | Medium anthropomorphic male (Male_Medium) | Medium anthropomorphic female (Female_Medium) | Low anthropomorphic genderless (Genderless_Low) |

Fig. 1. Five avatars in the online survey: High anthropomorphic male (Male_High), High anthropomorphic female (Female_High), Medium anthropomorphic male (Male_Medium), Medium anthropomorphic female (Female_Medum), and low anthropomorphic genderless (Genderless_Low)

*3.1.2 Scenarios Selection.* Blackwell et al. identified three categories of online harassment: verbal harassment, physical harassment, and environmental harassment [5]. This study specifically focuses on physical and environmental harassment. Guofreeman et al. pointed out the potential contradiction between users' avatar genders and the genders conveyed by their voices in social VR [28, 72]. The exclusion of verbal harassment from our focus serves to streamline the research process, as its consideration would introduce significant complexity. Furthermore, the inclusion of sexually suggestive language or other forms of verbal harassment in the survey questionnaire carries the risk of triggering content flags, potentially rendering the online survey inappropriate and subject to restriction.

In selecting specific behaviors, researchers referred to behaviors mentioned in previous studies [5, 28, 84]. For physical behaviors, we initially selected two behaviors: *Sexual Touching* and *Sexual Self-Touching*. Subsequently, we opted to examine two contentious behaviors: *Head Touching* and *Hugging*, since these behaviors are commonly observed in reality, but may be considered as harassment within VR. For environmental behaviors, we selected *Sexual Sketching*. Then, we combined 5 behaviors with 5 avatars. As a result, we obtained 25 scenarios and recorded them using VRChat (Table 1). All 25 scenarios appeared in each respondent's survey, but to ensure experimental accuracy, the order of scenario presentation was randomized to mitigate any order effects.

| Behavior / Avatar appearance | Head Touching | Hugging | Sexual Touching | Sexual Self-Touching | Sexual Sketching |
|---|---|---|---|---|---|
| High anthropomorphic male (Male_High) | 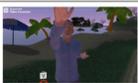 | 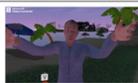 | 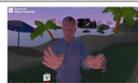 | 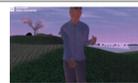 | 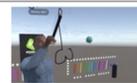 |
| High anthropomorphic female (Female_High) | 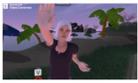 | 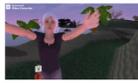 | 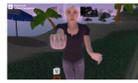 | 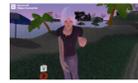 | 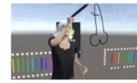 |
| Medium anthropomorphic male (Male_Medium) | 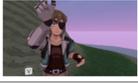 | 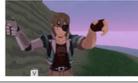 | 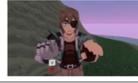 | 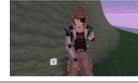 | 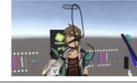 |
| Medium anthropomorphic male (Female_Medium) | 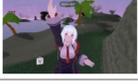 | 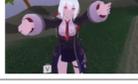 | 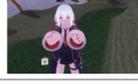 | 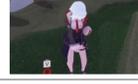 | 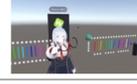 |
| Low anthropomorphic genderless (Genderless_Low) | 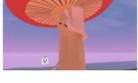 | 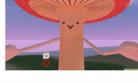 | 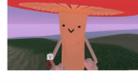 | 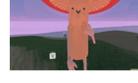 | 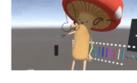 |

Table 1. 25 scenarios in the online survey ( 5 avatar appearances × 5 avatar behaviors).

*3.1.3 Questions design.* Figure 2 displays the survey questions for each scenario. Before participants answered the scenario-based questions, we provided a brief explanation for this section. For instance, we informed participants that they and the avatar were strangers and offered a concise interpretation of potentially unfamiliar terms. To understand users' perceptions, participants were instructed to rate whether the displayed scenario constituted sexual harassment on a 7-point Likert scale, with the opportunity to provide reasons. Regarding responses, a predefined set of choices was provided along with an "other" text box for participants to input additional answers: *NoRespond*, *Interact*, *HitBack*, *StayAway*, *SwitchAvatar*, *Report*, *Blocklist*, *SocialBoundary* and *CallPolice*. Here the *SocialBoundary* refers to the 'personal boundary/personal bubble space' provided by Horizon/Rec Room/VRChat. We provided the Chinese translation and a detailed description of this function in the questionnaire to ensure clarity. We initially drew upon the response strategies outlined by Guo et al. in their study on sexual harassment in social VR [28]. Additionally, researchers reviewed

VR harassment-related videos on Bilibili, one of China's most popular video platforms, and the comments that were written by users in the comments section below each video and summarized the response strategies to the behavior depicted in the videos. This analysis helped researchers identify the additional strategies that were not mentioned in prior studies.

Fig. 2. Example of a scenario question in the online questionnaire. A scenario question consists of a dynamic scene image and four accompanying questions.

## 3.2 Testing the Survey Tool

To test our survey, we conducted three stages of testing according to the method proposed by Dillman[16]. First, the survey was reviewed by experts in the VR field to uncover potential misunderstandings and potentially overlooked issues in the survey. Subsequently, we invited two bloggers who have published VR sexual harassment-related videos on Bilibili to fill out our survey and provide feedback on the topics, question design, and other aspects. This helped us ensure the content was well-motivated and the communication was effective. Finally, we performed pilot testing with four VR users to identify any flaws in the survey questions and distribution platform to ensure that the length was appropriate.

## 3.3 Respondent Recruitment

We used the "*Questionnaire Star*" platform (a popular questionnaire platform in China) to create and deliver the survey, which all respondents can access anonymously through a link. We extended the survey's reach by promoting it on social platforms like *WeChat*. In addition, as an incentive to take part in the survey, all respondents who recorded their email addresses could get early access to the project results and participate in a raffle for ¥100. Our study and recruitment were approved by the author's institutional review board (IRB).

## 3.4 Data Quality

After four weeks of data collection, we collected responses from VR users (N=171). To identify the invalid response, we referred to the category of inattentive respondents in Aaron Moss's method to detect speeders and straight-liners [58]. First, the median time for filling in the survey is 673 seconds. Respondents that took less than 30 of the median value (201.9s) are speeders [39]. Second, we calculated the standard deviation of respondents ' 7 Likert-scale answers for perceiving all 25 scenes and removed the ones with low standard deviation (stdev<0.2), which indicated straight-liners [43]. In summary, we found respondents P154 and P178 to be speeders and P60, P172, P178, and P275 to be straight-liners and got 166 valid responses.

Table 2. Demographic information of participants in semi-structured interview

| Participants | Gender | Age | VR platforms used |
| --- | --- | --- | --- |
| P1 | Male | 26-30 | VRChat, VRdesktop, ALYX, Beat Saber, VR Kanojo |
| P2 | Male | 21-25 | VRChat, VRdesktop, Beat Saber |
| P3 | Female | 21-25 | VRChat, Beat Saber |
| P4 | Male | 21-25 | REC Room, Beat Saber |
| P5 | Female | 26-30 | VRChat, Beat Saber |
| P6 | Female | 26-30 | VRChat, Beat Saber |
| P7 | Male | 21-25 | AltspaceVR, Beat Saber, VR Epic Roller Coasters |
| P8 | Female | 21-25 | REC Room, Beat Saber |
| P9 | Male | 21-25 | VRChat, REC Room, VR desktop, ALYX, Beat Saber, Blade and Sorcery, Cooking Simulator VR |
| P10 | Male | 21-25 | VRChat, VR desktop, ALYX, Beat Saber, Blade and Sorcery, Cooking Simulator VR |
| P11 | Male | 21-25 | VRChat, Beat Saber, Arizona Sunshine, Fruit Ninja |
| P12 | Male | 18-20 | REC Room, ALYX |
| P13 | Female | 21-25 | VRChat, Beat Saber, VR Kanojo |
| P14 | Female | 21-25 | AltspaceVR |
| P15 | Male | 21-25 | VRChat, Beat Saber |
| P16 | Male | 26-30 | VRChat, ALYX, Beat Saber |
| P17 | Male | 21-25 | REC Room, VRChat, ALYX |
| P18 | Female | 21-25 | VRChat, Beat Saber |

## 3.5 Respondent Demographics

We collected a total of 171 responses and the valid responses were 166. The median time to complete the survey was 680 seconds. Our respondents consisted of 80 males, 82 females, 2 others, and 2 who chose not to identify. The age of respondents who participated in the survey was mainly distributed between 21-25 years old ($N = 88$). Finally, the top three VR applications they have used are *VR Chat, Beat Saber, and VR Desktop.*

## 4 Semi-structured Interview

We conducted semi-structured interviews with volunteers from the survey respondents who were willing to participate in the follow-up interview. We used "Tencent Conference" for online interviews. All interviews lasted around 40 - 60 minutes and the interview video was recorded.

### 4.1 Participants

We conducted interviews with a total of 18 participants (denoted as P1-P18). Initially, we selected respondents who expressed their interest in participating in the interviews. We then reviewed their survey responses and chose participants who had prior experience using social VR applications and provided more extensive insights. We employed saturation as a criterion to determine the point at which data collection could be concluded. Hence, we ceased recruitment when no new topics or ideas emerged during the interviews. Our participant group consisted of 11 males and 7 females, with ages ranging from 18 to 30 years old. The VR applications they used included: 1) Social VR applications such as *VRChat* ($N = 13$), *REC Room* ($N = 5$), and *AltspaceVR* ($N = 2$); and 2) Other applications like *VRDesktop, ALYX, Beat Saber, Blade and Sorcery, Cooking Simulator VR, and more* (Table 2).

### 4.2 Procedure

Prior to the interviews, we summarized each participant's answers to the scenario questions in the online survey. We classified and categorized the responses into five figures based on behavior (Figure 3). This approach provided a visual representation, allowing us to examine how participants made judgments and responded when confronted with different avatars exhibiting the same behavior. During the interviews, these five figures were presented to participants to assist in recalling their initial answers.

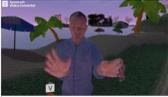

| Behavior 1 | 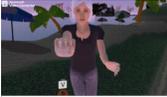 | 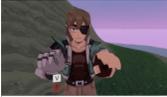 | 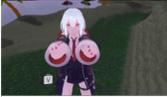 | 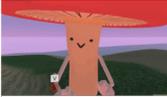 | |
| --- | --- | --- | --- | --- | --- |
| **What kind of behavior do you think the avatar is doing?** | Grope my chest | Grope my chest | Grope my chest | Grope my chest | Grope my chest |
| **This behavior is sexual harassment** | | | | | |
| **What is the reason for your rating?** | No answer | Touch private parts | No answer | She's adorable | No answer |
| **How would you respond to this behavior?** | | | | | |

Fig. 3. Interview materials example is given to participants: This figure depicts how respondents answered the four survey questions when confronted with five avatars displaying hugging behaviors.

At first, we asked participants about the reasons for their judgments on behavior and the factors that made different judgments despite the same behavior. For example, *"Why did you make this judgment about this scenario?"*, *"Did avatar appearance affect your judgment?"*. Secondly, regarding their response strategies, participants were asked to express the reasons for selecting specific response strategies and whether they encountered any difficulties in using those responses.

Finally, we asked participants if they had experienced harassment while using social VR, and (if so) how they judged and dealt with it.

### 4.3 Analysis Method

All interviews were conducted by two researchers in Mandarin as all our participants were Chinese native speakers. We recorded the entire interview using the meeting recording function of "Tencent Meeting".

Firstly, the two researchers thoroughly reviewed the interview transcript several times to have an overall understanding of sexual harassment in social VR. Then, they independently coded the script using an open-coding approach [79]. Employing both deductive and inductive coding techniques, they developed the codebook. Initially, two main preliminary themes were established: Behavioral perception and response strategies. Sub-themes and specific contents within each central theme were inductively constructed by assigning keywords to participants' feedback or answers. We grouped repeating keywords at a higher level and formed a multiple-level diagram. For example, when participants described the impact of avatar appearance on their behavioral perception, we labeled that section as the topic "Behavioral perception." When words such as "female," "friendly," and "non-aggressive" were repeated in the responses, we identified the sub-theme: "Female avatars are generally perceived as less aggressive and friendly." The two coders regularly discussed the codes and resolved disagreements to create a consolidated codebook. Additional meetings were scheduled with all co-authors to reach agreements based on the initial coding result. Finally, through three rounds of iterative coding, we generated two comprehensive themes: The appearance of the avatar moderates the user's behavioral perception, and users tend to use non-confrontational responses.

## 5 Results

In this section, we present the quantitative analysis results obtained from the survey in response to the research questions, along with qualitative analysis results obtained from semi-structured interviews.

### 5.1 Harassment Perception (RQ1)

*5.1.1 Quantitative Results.* We conducted statistical analyses on users' perception levels regarding whether the scenarios they face are harassing behaviors.

**Statistical method overview.** In this study, we employed mixed-effects regression analysis to investigate the **one-way, two-way, and three-way interactions** among the fixed effects, including the *potential harassing behaviors (e.g., head touching, hugging, sexually suggestive sketching, sexually self-touching, and sexually touching others)*, the *appearance of harasser's avatars (e.g., highly anthropomorphic male and female, moderately anthropomorphic male and female, and a control group of an avatar without any human or gender features)*, and the *gender of the participants (i.e., male and female)*. In our experimental design, we included the options *"Undetermined"* and *"Prefer not to disclose"* for non-binary participants to ensure the rigor of the experiment. However, in the data we obtained, only a very small number of individuals ($N = 4$) chose these options. Therefore, for quantitative analysis, we removed the data of these four participants (P113, P203, P284, P134). Participants were treated as random effects. We conducted all experiments on perception using the *nlme* package in R. In the following sections, we report the analysis result of how these three fixed effects impact users' perception levels of harassment and their potential interactions with each other.

**Overall performance measurements.** The mixed-effect regression model demonstrates that both behavior ($X^2(4)$=68.39, $p$<0.0001) and avatar appearance ($X^2(4)$=1952.00, $p$<0.0001) significantly influence user perception. However, the impact of participants' gender ($X^2(3)$=1.48, $p$=0.22) on perception does not exhibit significant differences across different genders. Figure 5 illustrates the distribution of user perception towards potential harassing scenarios based on one fixed effect only. The figure shows that users have a distinct perception regarding whether the behavior is harassing (*perception level* > 4). Behaviors such as touching the head and hugging are classified as non-harassing behaviors, while *SexualSelfTouching*, *SexualSketching* and *SexualTouching* are identified as harassing behaviors (Figure 4a). In addition, avatar appearance also significantly affects user harassment perception. Actions performed by male avatars are perceived as more harassing compared to those by female avatars. Avatars with a higher degree of anthropomorphism are also considered more harassing than those with a moderate degree of anthropomorphism (Figure 4b).

**Interactions between fixed effects.** The two-way interaction between avatar appearance and user gender exhibits a significant effect on harassment perception ($X^2(12)$=12.84, $p$=0.01<0.05), indicating that the level of perceived harassment towards avatars with different appearances varies among users of different genders. Figure 5a illustrates that overall, female users tend to perceive a higher level of harassment compared to male users. Specifically, male users feel the lowest level of harassment when facing female avatars, especially those with a moderate level of anthropomorphism. On the other hand, female users feel the highest level of harassment when facing highly anthropomorphic male avatars.

The two-way interaction between behavior and participant gender has a significant impact on user perception ($X^2(12)$=41.87, $p$<0.0001), indicating that there are differences in perception of harassing behaviors among users of different genders. Figure 5b illustrates the interaction between

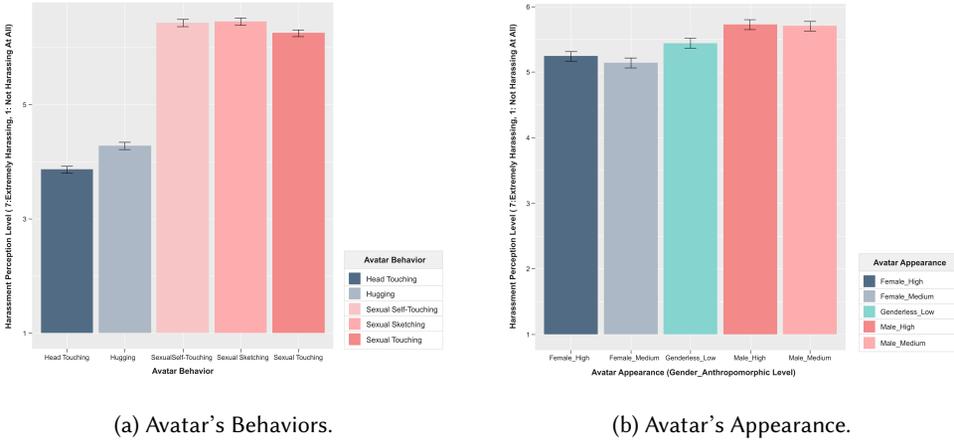

(a) Avatar's Behaviors.                                    (b) Avatar's Appearance.

Fig. 4.  Two-way interaction between avatar appearance and behavior on harassment perception. All error bars represent +/- standard error.

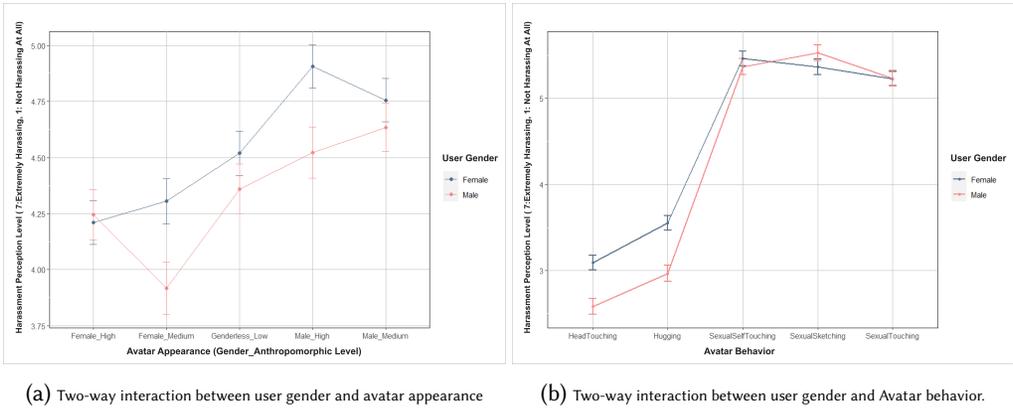

(a) Two-way interaction between user gender and avatar appearance          (b) Two-way interaction between user gender and Avatar behavior.

Fig. 5.  Two-way interaction between user gender and avatar behavior/ avatar appearance on harassment perception. All error bars represent +/- 95% confidence interval.

behavior and user gender, where even facing non-harassing behavior, female users still give higher perception scores. However, in the face of harassing behavior, male and female users have almost the same perception level.

The two-way interaction between behavior and avatar appearance also significantly influences user perception ($X^2$(16)=106.54, $p$<0.0001). This can be explained by the fact that users' perceptions of behavior are subject to change based on the appearance of the avatar. Figure 6 illustrates the interaction between behavior and avatar appearance. In Figure 6, we find that the perception level of users towards a female avatar is slightly lower than that towards a male avatar, regardless of the non-harassing or harassing behavior. Additionally, the impact of avatar appearance on user perception differs when facing different harassing behaviors. The perception level distribution corresponding to the *SexualSketching* is concentrated, indicating that the effect of appearance on perception is minimal when users encounter sexual drawing content. Conversely, the perception data for *SexualSelfTouching* and *SexualTouching* are relatively scattered. In *SexualSelfTouching*, the perception level associated with a highly anthropomorphic avatar is higher than that of a moderately anthropomorphic avatar. However, this phenomenon is reversed in *SexualTouching*.

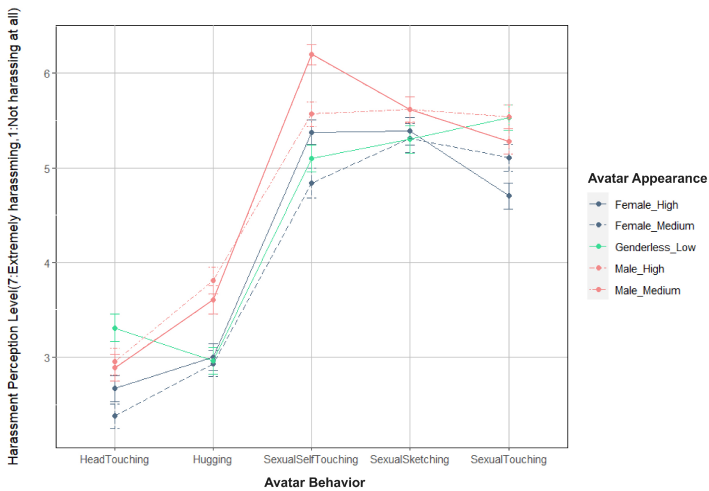

Fig. 6. Two-way interaction between avatar appearance and behavior on harassment perception. As shown by means and confidence interval. All error bars represent +/- 95% confidence interval.

We conducted a three-way interaction analysis on three fixed effects, and the results showed that the three-way interaction did not exhibit a significant effect on perception level ($X^2(48)$=14.86, $p$=0.53). This indicates that the interaction between avatar appearance and behavior does not differ among users of different genders.

### 5.1.2   Qualitative Results.
Data from subsubsection 5.1.1 indicates that while participants generally share a consistent qualitative understanding of sexual harassment behaviors, there are still variations in perception. Based on the interview data, it was observed that ***Avatar appearances play a moderating role in users' perception.*** This regulatory effect is primarily manifested in the following three aspects:

***The degree of anthropomorphism in avatars exerts varied influences on user perception.***
The data analysis of 5.1.1 reveals that avatars with a higher degree of anthropomorphism are perceived as more harassing than avatars with a moderate level of anthropomorphism. This phenomenon was elucidated during interviews. Many participants expressed unease when interacting with highly anthropomorphic avatars due to the heightened level of anthropomorphism, which intensified the feelings of immersion and realism, giving them the impression of being harassed by a "real person." Simultaneously, the limitations of VR devices maintain a distinction between highly anthropomorphic avatars and real individuals. This leads participants like P6 to *"feel discomfort stemming from the uncanny valley effect."* Another potential contributing factor is the varying prevalence of avatar types in VR communities. For instance, P1 and P9 observed a predominance of two-dimensional characters within the social VR community. Confronted with less common highly anthropomorphic avatars, they often demonstrate heightened vigilance.

Additionally, some participants expressed heightened discomfort when interacting with non-anthropomorphic avatars. They posited that users selecting less anthropomorphic avatars aimed to withhold personal information on social platforms, displayed an inability to establish genuine connections with others, and were perceived as untrustworthy. Therefore, encountering non-anthropomorphic avatars led to increased suspicion regarding users' true intentions, with speculation that their behavior was intentional. Moreover, P1 said:

> "...(This mushroom) itself is not a human, but when engaging in human-like behaviors, it appears obscene... I find the harassing actions performed by this avatar to be the most discomforting for me..."

**Female avatars are generally perceived as low-aggression and friendly.**

Many participants expressed that the avatar's gender played an important role in shaping their perception of harassment. Specifically, female avatars were consistently regarded as more amicable and less confrontational when juxtaposed with male avatars. Moreover, many participants correlated their perceptions with their real-world understanding of gender, such as men are commonly associated with physical strength, and women are often viewed as more susceptible to incidents of sexual harassment, as P8 said *"I believe that males are inherently stronger than females, and this perception is deeply rooted in our genes."* However, P8 did not entirely trust female avatars, as she was unable to determine whether the manipulator behind the female avatar was male.

> *"But I don't know who is under the VR headset. It could be a genuine beauty, or it might be someone entirely different, possibly a man, adopting an appealing facade to engage in peculiar actions toward you...In my previous gaming experiences, those who adopted attractive female avatars were mostly male."* (P8)

**Users' prior experiences shape a portion of their cognition of appearance and the perception of behavior.**

Participants acknowledged that their experience on social media, games, videos, and diverse platforms influenced their perceptions. Specifically, prior experiences infuse specific avatar appearances with novel connotations, thereby shaping the perception of behaviors linked to these appearances. Firstly, many participants correlated certain visual representations, including mushrooms, cucumbers, peaches, and eggplants, with genitalia and private parts of the human body, drawing insights from their interactions on social media and the internet. This results in some participants being more vigilant towards users employing such avatars, for P7, *"all actions performed by the mushroom avatar make me uncomfortable,"* as its appearance is deemed *"too suggestive."*

However, due to variations in cultural backgrounds and experiences, different users exhibit differences in their perceptions of the same avatar appearance. For instance, P11 described an observed difference in the use of emojis between Chinese and foreign users, he said,

> "I have observed that Chinese users tend to utilize mushrooms in online platforms, whereas foreigners may favor eggplants, and both groups may also employ cucumbers."

When faced with the same avatar, P17 remarked, *"I find this anime-style avatar very cute because I often encounter similar characters in Japanese animation."* In contrast, P2 directed attention to the black bands on the avatar's legs, linking them to specific cultures. P7, lacking familiarity with the cultural context behind this appearance, maintained a neutral stance toward the behavior associated with such avatars.

Additionally, many participants highlighted that their aesthetic preferences, molded by personal experiences, impact their perceptions. P8 expressed that the *"attractiveness"* of avatars holds sway over her, *"people are always more receptive to things that look good."* Nevertheless, delineating users' aesthetic standards remains highly subjective and personal, as our participants were unable to articulate clear and explicit criteria, as P3 said:

> "I find it challenging to articulate what I consider 'good-looking' since my assessments are often contingent upon my emotions and mood. Furthermore, these sentiments are subject to fluctuations and influenced by various uncertain factors."

## 5.2 Response Strategies (RQ2)

*5.2.1 Quantitative Results.* In this section, we conducted a frequency and distribution quantitative analysis of users' response choices from four perspectives: *potential harassing behavior, harasser avatar appearance, harassment perception level*, and *user gender*.

Table 3 shows the correspondence between the user's harassment perception with his choice of the response strategy. The darker the color of a table cell, the more participants have chosen that response option. This table indicates that responding strategies can be classified into two categories: **non-rejecting responses** (*NoRespond, Interact*) and **rejecting responses** (*HitBack, StayAway, SwitchAvatar, Report, Blocklist, SocialBoundary, CallPolice*). Users mostly choose non-rejecting responses when their perception level is between 1 and 3. However, when the perception level is 4, users are uncertain whether the behavior is harassment and thus tend to increase the number of people who choose to *StayAway* (41.17%). After confirming harassment, *StayAway, Report*, and *Blocklist* are the three most commonly used responding strategies. In addition, when facing extreme harassment (perception level=7), the number of people who choose to *HitBack* (44.07%) and establish a *SocialBoundary* (46.36%) also significantly increases, in addition to the previous three options.

Table 3. The contingency table of participants' responses to different levels of perceived harassment (in terms of percentage)

| Harassment Perception | NoRespond | Interact | HitBack | StayAway | SwitchAvatar | Report | Blocklist | SocialBoundary | CallPolice | Others |
|---|---|---|---|---|---|---|---|---|---|---|
| 1 | 455 | 264 | 12 | 28 | 14 | 5 | 9 | 12 | 0 | 5 |
| (Non-harassing at all) | (76.73%) | (44.52%) | (2.02%) | (4.72%) | (2.36%) | (0.84%) | (1.52%) | (2.02%) | (0.00%) | (0.84%) |
| 2 | 320 | 154 | 12 | 61 | 18 | 14 | 9 | 16 | 0 | 6 |
| | (76.74%) | (6.93%) | (2.88%) | (14.63%) | (4.32%) | (3.36%) | (2.16%) | (3.84%) | (0.00%) | (1.44%) |
| 3 | 215 | 85 | 25 | 82 | 11 | 19 | 20 | 21 | 2 | 6 |
| | (65.95%) | (26.07%) | (7.67%) | (25.15%) | (3.37%) | (5.83%) | (6.13%) | (6.44%) | (0.61%) | (1.84%) |
| 4 | 248 | 108 | 63 | 226 | 54 | 52 | 42 | 48 | 8 | 27 |
| | (45.6%) | (19.89%) | (11.60%) | (41.62%) | (9.94%) | (9.58%) | (7.73%) | (8.84%) | (1.47%) | (4.97%) |
| 5 | 94 | 82 | 131 | 370 | 113 | 215 | 177 | 126 | 20 | 8 |
| | (15.99%) | (13.95%) | (22.28%) | (62.93%) | (19.22%) | (36.56%) | (30.10%) | 21.43% | (3.40%) | (1.36%) |
| 6 | 31 | 69 | 180 | 431 | 174 | 365 | 334 | 201 | 21 | 3 |
| | (5.00%) | (11.13%) | (29.03%) | (69.52%) | (28.06%) | (58.87%) | (53.87%) | (32.42%) | (3.39%) | (0.48%) |
| 7 | 44 | 85 | 440 | 731 | 326 | 738 | 706 | 458 | 112 | 1 |
| (Extremely harassing) | (4.57%) | (8.83%) | (45.69%) | (75.91%) | (33.85%) | (76.64%) | (73.31%) | (47.56%) | (11.63%) | (0.10%) |

According to the distribution shown in Figure 4a between behaviors and perception, *HeadTouching* and *Hugging* are considered non-harassing behaviors, while *SexualSelfTouching, SexualSketching,* and *SexualTouching* are considered harassing behaviors. The results in the Table 4 show that for non-harassing behaviors, more users choose non-rejecting responses, but there are still many users who choose *StayAway* as a precautionary strategy. When faced with harassing behaviors, *StayAway, Report,* and *Blocklist* are still the most commonly selected response strategies.

Analyzing the data in Table 5, we find that users are more cautious around male avatars. Specifically, non-rejecting behavior *Interact* is more frequently selected in response to female avatars while rejecting behaviors are more frequently used in response to male avatars. The frequency of responses revealed that *StayAway, Report,* and adding the user to a *blocklist* are the most commonly chosen rejecting response choices, with over half of the users (Freq=52.53%) choosing to stay away from male avatars with a high degree of anthropomorphism.

In subsubsection 5.1.1, users' gender does not show significant differences in perception level, indicating that users' perception levels do not differ significantly across genders. The distribution of response strategy choices in Table 6 was similar between male and female users, with male users showing a slightly higher frequency of non-rejecting responses and female users more actively choosing rejecting responses.

Table 4. The contingency table of participants' responses to different behaviors (in terms of percentages=frequency/N, N = 162 participants x 5 behaviors)

| Harsser's behavior | NoRespond | Interact | HitBack | StayAway | SwitchAvatar | Report | Blocklist | SocialBoundary | CallPolice | Others |
|---|---|---|---|---|---|---|---|---|---|---|
| HeadTouching | 484 | 260 | 81 | 213 | 80 | 73 | 70 | 63 | 6 | 14 |
| | (59.75%) | (32.10%) | (10.00%) | (26.30%) | (9.88%) | (9.01%) | (8.64%) | (7.78%) | (0.74%) | (1.73%) |
| Hugging | 421 | 245 | 104 | 257 | 91 | 107 | 97 | 73 | 14 | 13 |
| | (51.98%) | (30.25%) | (12.84%) | (31.73%) | (11.23%) | (13.21%) | (11.98%) | (9.01%) | (1.73%) | (1.60%) |
| SexualSelfTouching | 159 | 112 | 220 | 487 | 168 | 421 | 397 | 248 | 54 | 10 |
| | (19.63%) | (13.83%) | (27.16%) | (60.12%) | (20.74%) | (51.98%) | (49.01%) | (30.62%) | (6.67%) | (1.23%) |
| SexualSketching | 187 | 85 | 210 | 487 | 177 | 441 | 393 | 263 | 55 | 10 |
| | (23.09%) | (10.49%) | (25.93%) | (60.12%) | (21.85%) | (54.44%) | (48.52%) | (32.47%) | (6.79%) | (1.23%) |
| SexualTouching | 156 | 145 | 248 | 485 | 194 | 366 | 340 | 235 | 34 | 9 |
| | (19.26%) | (17.90%) | (30.62%) | (59.88%) | (23.95%) | (45.19%) | (41.98%) | (29.01%) | (4.20%) | (1.11%) |

Table 5. The contingency table of participants' responses to different avatar appearances (in terms of percentages=frequency/N, N = 162 participants x 5 avatar appearances)

| Avatar Appearances | NoRespond | Interact | HitBack | StayAway | SwitchAvatar | Report | Blocklist | SocialBoundary | CallPolice | Others |
|---|---|---|---|---|---|---|---|---|---|---|
| Female-High | 289 | 198 | 147 | 366 | 133 | 232 | 201 | 159 | 22 | 12 |
| | (35.68%) | (24.44%) | (18.15%) | (45.19%) | (16.42%) | (28.64%) | (24.81%) | (19.63%) | (2.72%) | (1.48%) |
| Female-Medium | 293 | 230 | 155 | 343 | 133 | 247 | 236 | 158 | 28 | 11 |
| | (36.17%) | (28.40%) | (19.14%) | (42.35%) | (16.42%) | (30.49%) | (29.14%) | (19.51%) | (3.46%) | (1.36%) |
| Male-High | 265 | 133 | 195 | 431 | 159 | 326 | 314 | 200 | 49 | 9 |
| | (32.72%) | (16.42%) | (24.07%) | (53.21%) | (19.63%) | (40.25%) | (38.77%) | (24.69%) | (6.05%) | (1.11%) |
| Male-Medium | 269 | 138 | 191 | 406 | 150 | 313 | 290 | 179 | 31 | 12 |
| | (33.21%) | (17.04%) | (23.58%) | (50.12%) | (18.52%) | (38.64%) | (35.80%) | (22.10%) | (3.83%) | (1.48%) |
| Genderless-Low | 291 | 148 | 175 | 383 | 135 | 290 | 256 | 186 | 33 | 12 |
| | (35.93%) | (18.27%) | (21.60%) | (47.28%) | (16.67%) | (35.80%) | (31.60%) | (22.96%) | (4.07%) | (1.48%) |

Table 6. The contingency table of participants' responses to different user gender (in terms of percentage)

| User Gender | NoRespond | Interact | HitBack | StayAway | SwitchAvatar | Report | Blocklist | SocialBoundary | CallPolice | Others |
|---|---|---|---|---|---|---|---|---|---|---|
| Female | 620 | 312 | 440 | 1098 | 390 | 746 | 722 | 562 | 113 | 34 |
| | (31.00%) | (15.60%) | (22.00%) | (54.90%) | (19.50%) | (37.30%) | (36.10%) | (28.10%) | (5.65%) | (1.70%) |
| Male | 787 | 535 | 423 | 831 | 320 | 662 | 575 | 320 | 50 | 22 |
| | (38.39%) | (26.10%) | (20.63%) | (40.54%) | (15.61%) | (32.29%) | (28.05%) | (15.61%) | (2.44%) | (1.07%) |

In addition, *CallPolice* has the lowest frequency of selection in the frequency tables 3 to 6, suggesting that users make a clear distinction between reality and virtual reality. Harassment incidents that occur in social VR environments should be addressed within the virtual space.

In summary, no significant effect is found in users' perception levels across different genders, and users' judgments are largely based on behavior. Female users exhibit slightly higher perception levels when facing non-harassing behaviors. Avatar appearance has a secondary impact on harassment perception, and its influence varies depending on the behavior. In our study, the impact of avatar appearance is the smallest in *SexualTouching*, and avatar's anthropomorphic level exhibits diverse effects on *SexualTouching* and *SexualSelfTouching*. When male and female avatars engage in the same behavior, the actions of the male avatars are perceived as more severe forms of harassment.

Depending on the scenario, response strategies can be divided into non-rejecting responses and rejecting responses. *StayAway*, *Report*, and *BlockList* are the three most commonly used rejecting response strategies. In addition, *StayAway* is frequently chosen even in low-perception scenarios. *HitBack* and establishing the *SocialBoundary* are also adopted by users when facing highly perceived harassing behavior. Female users are more active in using rejecting responses to protect themselves compared to male users. Besides, users have no demand that harassers on social VR platforms should face real-life consequences for their actions.

*5.2.2   Quantitative Results.* Combining the quantitative analysis results, we found that users tend to adopt non-confrontational responses (*stay away, report, blocklist*). Therefore, through the interviews, we identified several factors that contribute to this phenomenon.

**Non-confrontational response strategy is considered to be cost-effective.**

At first, some participants perceived the primary functions of social VR platforms as sources of entertainment and relaxation. They believed that harassment in social VR did not directly threaten their personal property. Consequently, they exhibited a preference for non-confrontational approaches to uphold a positive social VR experience. Simultaneously, adopting a non-confrontational approach allows certain participants to promptly extricate themselves from the dilemma without compromising their interests. For instance, P8 and P14 expressed their rationales for staying away instead of changing avatars.

> *"If they know you're a woman, changing the avatar may not deter harassment entirely. Beyond physical gestures, it may persist in textual harassment and similar forms. Therefore, I tend to stay away from them, severing any potential for further influence."* (P8)

> *"I feel compelled to create distance...switching the avatar becomes a minor sacrifice for me. Why should I alter the avatar I favor due to harassment?"* (P14)

Additionally, given the ambiguity associated with certain behaviors, users often adopted a cautious "stay away" approach to mitigate potential misunderstandings. As expressed by P7,

> *"I don't comprehend the nuances of anime culture, so I am uncertain about the expressions, and these actions might be their means of friendliness... I tend to maintain a distance from these avatars."*

However, for P6, encountering harassment while playing with friends, "reporting" and "blocking list" hold more tangible benefits than opting for "staying away."

> *"If I am with friends, choosing to distance myself would imply everyone leaving together. I find this a more challenging approach. Comparatively, if he is intentionally harassing me, I would opt to report or block him..."*

Moreover, reporting was not universally regarded as effective. For instance, P5's experience with various online games led her to express concern about the promptness of feedback after reporting, and she stated that the *"reporting results were untimely and not always satisfactory."* P13 felt that there was still a possibility of encountering harassers after reporting. In contrast, she believed that *"blacklisting could prevent any future encounters with harassers."*

**Users will adopt response strategies corresponding to the progression of behaviors.**

We observed that certain users frequently adjust their response strategy based on the stage of the situation and the feedback received after implementing an action. This adaptive approach, employed by this subset of participants, serves to mitigate conflicts arising from misunderstandings of behavior and minimizes the time invested in coping. For instance, P1 initially categorizes some harassing behavior as pranks, displaying a degree of tolerance and attempting communication with the other party. In cases where communication proves ineffective, P1 *"employed strategies such as reporting, allowing the system to adjudicate or directly blocking the offender."*

Additionally, P6 indicated that she tends to opt for staying away. If the other party does not persistently follow, P6 refrains from taking further action. However, if the other party persists in continuous harassment, P6 stated,

> *"If I enjoy the current environment and someone keeps following me and engaging in such behavior, in that case, I would resort to reporting, blocking, or even hitting back!"*

***Users' response decisions would be influenced by the presence or absence of bystanders.***
Some participants choose confrontational responses in the presence of bystanders, as they believe that bystanders can bolster their courage to counteract, intensify the harasser's sense of shame, and act as witnesses to substantiate their case. As P5 said:

> *"In public chat rooms, comment sections, or public spaces, I typically adopt the 'Hit Back' approach...I vividly recall someone advising, 'Speak up when you encounter harassment on a bus because there will be many people to support you.' Consequently, when confronted with harassment in social VR, I also take into account the presence of bystanders."*

P18 agreed, underscoring the significance of bystanders, particularly friends, in encouraging her confidence to respond assertively. She expressed, *"To be honest, I feel apprehensive about engaging in arguments with others, both in real life and VR. However, with the support of my friends, I may summon the courage to intervene and halt the harassment."*

## 6  Discussion

We adopted a mixed-methods approach to investigate the effects of the harasser's avatar appearance and behavior on users' perception of sexual harassment on social VR platforms and how users tend to respond when encountering such sexual harassment. As prior studies were primarily conducted in Western countries, our work contributed to the limited understanding of Chinese VR users on harassment on social VR platforms.

In this section, we first highlight key findings based on the results of our quantitative and qualitative analyses and discuss potential reasons behind our findings. Furthermore, we discuss how our findings extend the existing literature in the CSCW and HCI communities. We present new features of how users perceive harassment on social VR platforms and the support and barriers they face when choosing how to respond to sexual harassment. Finally, we discuss the opportunities and challenges that exist in data regulation based on our findings from semi-structured interviews in conjunction with the existing literature on social VR platform data.

### 6.1  *RQ1*: Factors affecting harassment perception

Our findings indicate that users exhibit consistent judgments regarding sexual harassment, despite the subjective nature of harassment definitions as noted in [5]. This suggests that social VR communities may be capable of regulating sexual harassment. Furthermore, we found that user gender does not significantly impact their perception of potential sexual harassment behaviors. A significant body of literature has explored how social VR users perceive avatars[5, 26, 28], highlighting the vulnerability of certain avatar features, such as those associated with marginalized groups and females. However, these studies have primarily focused on the impact of the victim's avatar.

Our study extends the research on a single variable takes behavior into consideration and finds that avatar appearances exhibit different effects on different behaviors. Figure 6 shows that users are less influenced by avatar appearance when perceiving *SexualSketching*. For the other two harassing behaviors, male avatars exhibit higher levels of harassment than female avatars. However, the impact of avatar anthropomorphism differs between these two behaviors. Higher levels of avatar anthropomorphism led to greater perceived harassment in *SexualSelfTouching*, while less harassing in *SexualTouching*.

#### 6.1.1  *Perception as combination of focus area and focus level.*  In our study, we found that users exhibit varying levels of effect on the appearance of harassers when faced with different types of harassment. We propose that the overall perception of users is a combination of *focus area* and *focus level*.

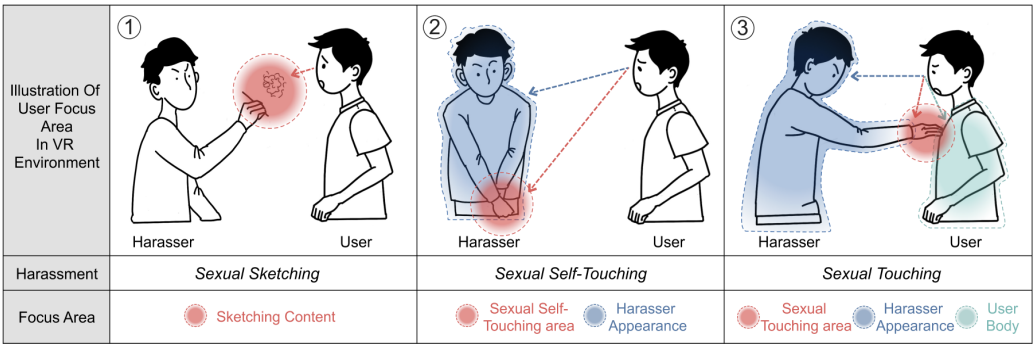

Fig. 7. Different types of focus area for *SexualSketching*, *SexualSelfTouching*, and *SexualTouching* in VR environment

When confronted with *SexualSketching*, users primarily focus on the created content (red area on Figure 7), with a low focus level on the harasser's appearance. In the case of *SexualSelfTouching*, users initially focus on the actions performed by the harasser (red area on Figure 7), but inevitably pay more attention to the harasser's appearance (blue area on Figure 7). In this situation, the avatar's appearance can convey more information and have a greater impact.

*SexualTouching* is a more complex scenario where, in addition to attention given to the harasser's behavior (red) and appearance (blue), there is a strong target focus (green). Moreover, besides visual influence, there is *visuotactile* perception, as demonstrated by Kammers et al. through the "rubber hand" experiment [41], indicating that visual stimuli can evoke tactile sensations. This phenomenon is enhanced in virtual reality environments, allowing users to experience virtual touch more realistically. This visuotactile also diverts the user's attention from the harasser's appearance.

Therefore, in the aforementioned three harassment scenarios, each could have multiple focus areas, the order of focus levels on avatar appearance is as follows: *SexualSelfTouching* > *Sexual-Touching* > *SexualSketching*, which aligns with the extent to which appearance influences each scenario as observed in the data.

In conclusion, we assert the concepts of focus area and focus level can aid community moderators in comprehending the diverse influence of users' avatar appearances across different instances of harassment. It facilitates the development of more precise user avatar designs by leveraging the interactive capabilities of social platforms. Additionally, it offers deeper insights into the nature of harassing behaviors, enabling the evaluation of whether a user's avatar contributes to the factors involved in harassment.

*6.1.2 Dichotomy effect of avatar anthropomorphism.* Our findings demonstrate that users pay little attention to the avatar appearance of the harasser when encountering sexual sketching. However, in the case of *SexualTouching* and *SexualSelfTouching*, users' perception of the harasser's avatar appearance exhibits contrasting results. Drawing from prior studies, we learned that users in VR tend to have a preference for highly anthropomorphic avatars. The reason behind this preference is that realistic textures make the avatar more relatable and trustworthy, thus users prefer realistic avatars in social interactions and collaboration [40, 46, 86]. However, research has also highlighted the counter-effects of realistic avatars, such as increased anxiety among users during public speaking [31] or language learning [49]. These findings align with our experimental results, when engaging in harassing behaviors, users experience a greater sense of discomfort and a stronger resistance as highly anthropomorphic avatars are more trusted by users in social VR. It is worth mentioning that in our results, the avatar with the low anthropomorphic genderless avatar (Mushroom) exhibited higher levels of harassment compared to Medium anthropomorphic female avatar. We discuss the potential reason for this outlier in Section 6.1.3.

However, this phenomenon does not persist in the context of *SexualTouching*. In the interview, participants manifested their concerns regarding individuals concealing their true identities using non-human avatars. Some users intentionally exploit this anonymity to harass others. Hence, when users are approached by less anthropomorphic embodiments, they feel more uncomfortable because these individuals subjectively hide their "human" attributes, thereby reinforcing the purposefulness of the harassment. Similarly, other researchers have investigated the impact of avatar appearance on pre-touch proxemics in VR environments. They found that, compared to human-like avatars, users tend to maintain a greater distance from anthropomorphic robot avatars before being touched [44]. Similar conclusions were drawn in [38], where the study introduced a lower degree of anthropomorphism in the form of a cylinder, resulting in increased distance. Huang et al.'s research also revealed that users prefer human-like agents over pillar agents [37]. Our findings demonstrate that avatars with the lowest level of anthropomorphism exhibit the highest level of harassing in *SexualTouching* scenario, which is consistent with previous research.

Our study discovered that the degree of anthropomorphism of avatars has a distinct influence across different harassment scenarios. This discovery can assist administrators of social VR platforms in developing more refined regulatory measures for avatars with varying degrees of anthropomorphism. They can establish limitations on the degree of anthropomorphism based on the platform's positioning, aiming to strike a balance between user freedom of self-expression and prevention of harassing behaviors.

### 6.1.3 Effect of Prior Experience and Culture on Avatar Appearance Cognition. Our study found that users form perceptions of specific avatar appearances based on personal experiences, which subsequently influence their understanding of the behaviors exhibited by these avatars. Additionally, users' perceptions of appearance are primarily influenced by online platforms such as social media and games. For example, certain emoji convey flirtatious information [35, 71], such as interpreting the "eggplant" emoji as male reproductive organs and the "peach" emoji as buttocks [57, 78, 81]. Furthermore, participants integrated pre-existing gender-related cognitive notions and stereotypes from the real world into their perceptions of appearance and behavior. For instance, when engaging in similar behaviors, the behaviors of male harassers are more frequently interpreted as sexual harassment, aligning with findings in other pertinent studies [23, 45].

In conclusion, given that the primary identifier for users in the virtual world is the avatar, and considering the high level of avatar customization [18], users are likely to evaluate avatar behaviors according to prior cognitive notions around appearance. However, there exist distinct individual differences in prior experience. For instance, despite our users sharing a prior understanding of certain emojis, similarities, and differences in emojis usage across platforms and regions [34, 42] may result in inconsistent interpretations among users, impacting their perception of the same avatar. Concurrently, prior experience is closely tied to culture, while the relationship between avatar types and culture is closed [85]. For example, East Asians tend to favor non-realistic and cartoon-like avatars, whereas Americans typically prefer realistic and human-like avatars [53, 85].

The above discussion from literature can explain the outlier observed in section 6.1.2 to some extent: mushrooms were perceived as more harassing compared to cartoon female avatars. Prior experience, as well as users' preferences for avatar selection in different cultural backgrounds, can explain the influence on interpreting avatar appearance: when users pay greater attention to avatar appearance, the experience of mushrooms as a potentially suggestive symbol influences user's perception. Moreover, in the context of East Asia, users tend to be more accepting of cartoon avatars, particularly those depicting female characters.

In sum, the significant differences in prior experience arising from either platform disparities or cultural variations greatly heighten the complexity of the platform's behavior regulation. Therefore, considering that avatar appearances on most social VR platforms are currently provided by the platform and limited user editing capabilities, we propose the following recommendations: 1) Platforms should evaluate the universality of avatars during the initial design, actively avoiding controversial avatar appearances. For platforms permitting user uploads and designs, like VRChat, it is imperative to provide upfront information about potential conflicts arising from sensitive images and features; 2) Cultural preferences and appropriateness [2] should be considered in avatar design, behavior supervision, and management within various cultural communities. Reasonable judgments should be made with a comprehensive understanding of the respective culture.

## 6.2  *RQ2* : Preference of respond strategies

We conclude similar results to literature [5, 26, 28] that females as an approachable but vulnerable group. Users are more cautious with male avatars. Moreover, female users are more inclined to utilize rejecting responses as a means of self-protection.

Among the rejecting responses, *StayAway*, *Report*, and *BlockList* are the most popular, while *HitBack* and *SocialBoundary* are comparatively less favored. *StayAway* is the preferred response strategy in Social VR due to its high efficiency, lack of restrictions, and freedom of operation, which allows users to leave the room at any time and is favored even in non-harassment situations, providing protection unavailable in real-life situations. *Report* and *BlockList* are popular due to their widespread use on social media platforms. Prior research also mention these response strategies, noting that *StayAway* is a personal technique, while *Report* and *BlockList* are platform-specific tools [28, 72]. It also indicates *BlockList* is the most effective way to prevent interactions with a blocked user, while *Report* as retroactive strategy to the platform.

Guo Freeman claims that  the creation of a personal space bubble is praised as highly suitable for Social VR platforms. However, our findings suggest that the establishment of social boundaries is not as popular as expected. We hypothesize that *SocialBoundary* is a mechanism unique to Social VR platforms, and users are not familiar with it. Meanwhile, showing social boundaries in front of the harasser is described as *confrontational* by some interviewees.

It is noteworthy that, in the discussion of subsection 6.1.3, we explicitly described the impact of prior experience on perception. Similarly, response strategies have been observed to be influenced by these prior experiences. On one hand, users' online experiences affect their responses. Harassment on online platforms is not uncommon, however, according to the Pew Research Center, 60% of males and females choose to ignore online harassment [65]. This phenomenon is also observed in Blackwell's study on social VR, where users perceive harassment as commonplace in online gaming, leading to desensitization to harassment in social VR [5]. Users are less inclined to respond aggressively to these behaviors. This partially explains why *Stay away* becomes a highly accepted response. Additionally, other related studies have found that the reluctance to respond may be due to inadequate online platform supervision and handling mechanisms [24], this experience may affect users' trust in the anti-harassment capabilities in social VR. On the other hand, users' real-life experiences influence their responses, as in work scenarios, individuals facing harassment often attempt to disengage from the environment as quickly as possible [30], aligning with the highly accepted response in social VR. Furthermore, the observed support effects for bystanders are also reliant on users' real-life experiences to some extent [54, 72], this discussion we will elaborate on in detail in subsubsection 6.2.2.

*6.2.1 'Confrontational' as further interaction with 'harasser'.* The findings in [28, 72] suggest that users are often unwilling to use platform-specific tools because these tools are too *confrontational*. We found that users regard *confrontational* as whether or not they have to interact with the perpetrator when executing a response strategy. *StayAway*, *Report*, and *BlockList* are all response strategies that users can operate independently; a harasser who has been blocked or reported cannot initiate interactions with the victim. In contrast, establishing a *SocialBoundary* and *HitBack* are more confrontational, because they involve visualized boundary or further physical interaction. Similar to real-life situations, users in Social VR platforms are facing avatars rather than just profile pictures, some users tend to prefer non-confrontational response strategies. We agree with the conclusions of previous studies and have supplemented the definition of confrontational as the requirement of interacting with the harasser in their awareness. Additionally, we suggest that future research explore strategies that assist some users in taking "confrontational" actions to protect themselves without engaging in frequent interactions with harassers.

*6.2.2 Supportive Effect from Presence of Bystanders.* Schulenberg mentioned that the company of friends can support female users in confronting harassers [72]. Besides friends, here in our study, interviewees also mentioned the presence of bystanders can also support their choice of confrontational response strategies. Previous studies have shown that bystanders play a crucial role in sexual harassment incidents, and a proactive bystander can provide support to the victim and encourage them to fight back against harassment [6, 20, 47, 59]. However, these studies also show that bystanders have diverse profiles, and pluralistic ignorance and diffusion of responsibility [15] may prevent them from making prompt and positive reactions. Blackwell [5] suggests that the immersive nature of Social VR platforms can enhance users' empathy and sense of responsibility. Owing to such enhancement and reduced risk, we hypothesize that users in Social VR exhibit a great potential to assume the role of proactive bystanders.

## 6.3 Opportunities and Concerns for Multi-modal Information Moderation

The significance of the *Report* function as a platform-specific tool in Social VR platforms has been emphasized in [28, 72]. As a retroactive feature, the report function enables users to access the platform to punish harassers after experiencing harassment. Our quantitative analysis reveals a large number of users who choose to report, but they also acknowledge that the report function was mostly ineffective. We speculate that there may be three reasons behind the ineffectiveness: (1) the **instantaneous nature of events in Social VR**, (2) the **trade-off between data regulation and privacy protection**, and (3) the **complexity of multi-modal data regulation**.

Our research, as well as the literature [5, 28, 72], highlights that the reason why harassment in Social VR is far more severe than on other online platforms is due to its high level of immersion and violation of physical space. This violation often occurs through instantaneous actions, unlike social media or online games where voice or text messages can be archived. Such instantaneous nature makes the retroactive report function fundamentally difficult to implement.

To combat instantaneous harassment in Social VR, recording everything in the scene is an obvious solution, such function is also recommended and researched in the design of VR collaboration platforms [10]. But this raises privacy concerns as recording can be detailed down to every joint in the user's avatar. The misuse of this data poses significant risks, hindering social VR platform growth. Social VR platforms should establish appropriate boundaries for data collection, disclose such boundaries to users, and obtain their consent. Schulenberg et al. suggest that AI moderators have the potential to provide significant assistance in community regulation for social VR platforms. They proposes that user-human-AI collaboration may effectively leverage recorded platform data, engage in mutual supervision to further enhance regulatory efforts while ensuring transparency and controllability of technology [73].

In addition, the issue of multi-modal data for regulation should also be considered. Literature [5, 26, 28, 72] have shown the diverse types of data that can be collected in Social VR platforms, such as avatar appearance and movement, spatial information, voice, and in our study, we propose that victim's focus area and focus level  and bystander presence is also a valid data for harassment detection, all of which can be used for retrospective analysis of sexual harassment behavior. Therefore, the collection, processing, and interpretation of multi-modal data are all highly complex tasks.

In summary, Social VR surpasses social media in terms of immersion and spatial perception, making it more like the real world. However, each user exists on the platform in the form of data. Theoretically, the platform is capable of monitoring every action performed by users on the Social VR platform. From a technical perspective, researching how to detect or even prevent instances of harassment in a short time, while also preventing data overload and privacy infringement caused by excessive data storage, is a critical research direction. AI moderators are likely to become important aids for regulating Social VR platforms in the future. Schulenberg et al.'s work also underscores the significance of user-human-AI collaboration in the regulation of harassment incidents within social VR platforms [74]. Different AI models can solve the difficulties of multi-modal data regulation. Determining how to correctly allocate weights to different types of data to obtain more accurate results, and how to achieve user-human-AI cooperation to ensure model interpretability, controllability, and transparency, are also research directions that deserve attention.

## 7 Limitation and Future Work

Firstly, our study focused on investigating the effects of male and female avatars on user perception and response strategies, aiming to control experimental variables and concentrate on specific research directions. However, with the increasing popularity of VR and social VR, the diversity of user gender on platforms has become more prominent. In the future, there is a need for further exploration to understand how users perceive and respond to harassment behaviors when facing other gender avatars. Secondly, the participants recruited for this study were users of popular commercial social VR platforms such as VRchat, REC room, and Altspace. The study did not encompass a broader range of users from various social VR platforms. Therefore, future work should aim to recruit samples from different social VR platforms to gather diverse user perspectives. Third, this study focuses on Chinese users, further research should supplement with a more diverse range of cultural backgrounds to understand the similarities and differences in the behavior of users from different cultural backgrounds in VR social (such as the interpretation of the mushroom avatar). Finally, we discover that users' prior knowledge influences the interpretation of avatar appearance. This finding can serve as a foundation for future research directions in the field of avatar interpretation.

## 8 Conclusion

In recent years, there has been an increasing number of reported cases of sexual harassment in social VR. Therefore, we employed a mixed-methods approach (survey + semi-structured interviews) to explore the influence of avatar appearance on user perception of behavior and their tendencies in response strategies. We found that users' perception of harassment behavior is consistent but modulated by avatar appearance, users tend to adopt non-confrontational response strategies. Based on these findings, we conducted in-depth discussions. Firstly, we explored the reasons why avatar appearance can modulate behavior perception, including attention-based information transformation the dichotomy effect of avatar anthropomorphism, and the influence of previous experiences. Secondly, we discussed the reasons why users tend to adopt "non-confrontational" responses and analyzed the positive role of "bystanders." Lastly, we highlighted opportunities and concerns for Multi-modal information moderation. Our study extended the existing literature in a new perspective of the harasser's avatar, addressed new features for perceiving and responding to sexual harassment, and proposed future research directions for multi-modal data regulation on social VR platforms.

## Acknowledgments

This work is partially supported by Guangzhou-HKUST(GZ) Joint Funding Project (No.: 2024A03J0617), the Guangzhou Municipal Nansha District Science and Technology Bureau under Contract (No.: 2022ZD012), CCF-Lenovo Blue Ocean Research Fund (No.: CCF-Lenovo OF 202302), HKUST Practice Research with Project title "RBM talent cultivation Exploration" (No.: HKUST(GZ)-ROP2023030), and Guangzhou Science and Technology Program City-University Joint Funding Project (No.: 2023A03J0001). In addition, thanks to Aoyu Wu for his invaluable guidance and assistance throughout this study. We also thank all our participants for taking the time to share their feedback.